\documentclass[aps,amsmath,amssymb,twocolumn,showpacs,longbibliography]{revtex4-1}

\usepackage{graphicx}
\usepackage{dcolumn}
\usepackage{bm}
\usepackage{color}

\begin{document}
\title{Variability of the thermal conductance of gold-alkane-gold single-molecule junctions studied using ab-initio and molecular dynamics approaches}
\author{J. C. Kl\"{o}ckner$^{1,2}$}
\author{F. Pauly$^{1,2}$}
\email{Fabian.Pauly@oist.jp}
\affiliation{$^{1}$Okinawa Institute of Science and Technology Graduate University, Onna-son, Okinawa 904-0495, Japan}
\affiliation{$^{2}$Department of Physics, University of Konstanz, D-78457 Konstanz, Germany}

\begin{abstract}
  Motivated by recent experiments, we study the variability of the thermal
  conductance of single dithiolated alkane molecules of varying length
  connecting two gold electrodes. For this purpose, we examine (i) the
  influence of the metal-molecule contact and of the electrode orientation on
  the thermal conductance of straight alkane chains, (ii) the effect of
  molecule-internal disorder realized through torsional gauche defects and the
  behavior upon stretching, and (iii) the modifications resulting from
  temperature-dependent dynamical variations of the geometry. While we analyze
  the former two aspects with a combination of density functional theory (DFT)
  and nonequilibrium Green's function (NEGF) methods, in the latter case we
  use nonequilibrium molecular dynamics (NEMD). Our calculations show the size
  of the variation of the phonon thermal conductance due to changes in contact
  geometry, that gauche defects generally reduce the thermal conductance but
  that they vanish upon stretching of the junction, and that the thermal
  conductance at elevated temperatures results from an average over many
  junction configurations, including alkanes with thermally excited gauche
  defects. Overall we find a good agreement between the DFT-NEGF and NEMD
  approaches for our room-temperature investigations. This confirms that
  anharmonic effects do not play a major role in alkane chains containing
  between four to ten carbon atoms. The phonon heat transport thus proceeds
  elastically and phase-coherently through these short molecular junctions.
\end{abstract}

\maketitle
\section{Introduction}
Since the pioneering theoretical proposal of a rectifying single-molecule
device \cite{Aviram1974} and the first charge transport measurements on single
molecules \cite{Reed1997,Smit2002}, many groundbreaking experiments have
explored fundamental physical aspects in these kinds of nanojunctions. They
involve the electrical conductance \cite{Nitzan2003,Xu2003}, thermoelectricity
\cite{Reddy2007,Cui2018}, energy dissipation \cite{Lee2013}, quantum
interference \cite{Garner2018,Miao2018} and electronic noise
\cite{Djukic2006,Karimi2016,Lumbroso2018}. All of these studies contributed
significantly to our current understanding of the phase-coherent electronic
charge transport at the nanoscale, thereby establishing the active research
field of molecular electronics \cite{Baldea2016, Cuevas2017,
  Moth-Poulsen2016}. One of the most appealing features of molecular junctions
is their tunability in terms of chemical synthesis, including molecular length
\cite{Hines2010}, conjugation \cite{Venkataraman2006,Mishchenko2010},
anchoring groups \cite{Park2007,Leary2015}, and chemical substituents
\cite{Venkataraman2007}, as well as the possibility to exert external control,
for instance through mechanical forces \cite{Vacek2015, Stefani2018} or
photons to trigger optically induced switching \cite{Dulic2003}.

Only recently the molecular junctions have been pushed into an entirely new
direction by measuring their thermal conductance \cite{Cui2019}. Since most
molecules exhibit a rather insulating, off-resonant charge transport behavior,
electronic contributions to the thermal conductance can typically be
disregarded \cite{Kloeckner2016,Kloeckner2017,Cui2019}. The thermal
conductance thus arises basically from phonons or photons
\cite{Kloeckner2017a}. Due to the recently established measurement scheme of
self-breaking single-molecule junctions \cite{Cui2019}, which will be
described further below, radiative heat transport plays no role, and the
phononic thermal conductance contribution is directly determined. This allows
to explore the physics of bosonic quasiparticles at the nanoscale. Similar to
the field of ``molecular electronics'' the upcoming field of ``molecular
nanophononics'' will benefit from the molecular tunability. Theoretical
suggestions have already been made to realize interference effects or to
adjust the thermal transport by use of anchoring groups and substituents
\cite{Markussen2013,Kloeckner2017,Kloeckner2016,Li2017,Famili2017}. With the
novel measurement techniques \cite{Cui2019}, the theoretical predictions are
now amenable to the experimental test.

The thermal transport through self-assembled monolayers of alkane molecules
has been studied by several research groups \cite{Wang2006, Wang2007,
  Meier2014}. Genuine single-molecule experiments eliminate crucial
uncertainties of such ensemble measurements, for example the number of
molecules contacted, the role of intermolecular interactions, possible defects
in the self-assembled monolayers or thickness variations due to surface
corrugation. Ref.~\cite{Cui2019} has successfully demonstrated
room-temperature measurements of the thermal conductance of gold-alkane-gold
single-molecule contacts. By detecting a length-independent thermal
conductance for dithiolated alkanes of different lengths at room temperature,
a basic fingerprint of the phase-coherent nature of heat transport in these
nanojunctions has been shown.

The key experimental challenge that has been overcome in Ref.~\cite{Cui2019}
is the detection of minute thermal currents through single-molecule junctions
at room temperature. Depending on the applied temperature difference, thermal
currents are on the order of picowatts, and a highly sensitive thermal
measurement scheme is needed that additionally ensures a high mechanical
device stability at the elevated temperatures. In order to measure the thermal
conductance at room temperature, the following advanced protocol has been
developed: In a first step, molecular junctions are characterized
electronically through a conventional electrical conductance histogram. The
information contained in the histogram is then used to identify
single-molecule junctions through the most probable single-molecule electrical
conductance. Electrical conductance-distance traces are now taken, but the tip
withdrawal is stopped, if a single-molecule junction is identified. At this
point the electrical conductance and the temperature of the tip, from which
the thermal conductance of the junction can be deduced, are monitored
simultaneously. After a waiting time on the order of seconds, the contact
finally breaks spontaneously. Due to the low thermal signal, a sophisticated
averaging scheme needs to be applied. Several hundred thermal conductance-time
traces, each 1.1~s in length, are aligned to the breaking point, identified by
the electrical conductance signal, and averaged. In this way a clear change in
the thermal conductance due to contact rupture is observed, which is identical
to the sum of the phononic and often negligible electronic thermal conductance
of the single-molecule junction.

While it would be desirable to measure a full thermal conductance-distance
trace, similar to what is possible in electrical conductance measurements,
with the newly developed protocol the single-molecule thermal conductance can
only be studied at the point of breaking. On the other hand, it is a clear
advantage that the measurement protocol determines the thermal conductance as
a difference of thermal conductances before and after the thermally induced
contact rupture. Since the electrode separation remains unchanged, radiative
thermal conductance contributions cancel out. In Ref.~\cite{Cui2019} the
alkane chains C2-C10, containing two to ten carbon atoms, have been
analyzed. For C2 the electronic contribution to the thermal conductance was
significant, but it could be neglected for C4-C10. For these reasons we focus
here on phonon thermal conductance calculations for C4-C10. As the experiment
determines the phononic thermal conductance as a mean value over a time series
of junction structures before contact rupture, it is important to
theoretically explore the magnitude of geometry-induced variations of the
thermal conductance. We address this topic in the present work.

Our study is structured as follows. We commence with a description of the
theoretical methods in Sec.~\ref{sec-theory}, where we introduce the DFT-NEGF
method for phonon transport in Sec.~\ref{sec-theory-DFT+NEGF} and the NEMD
approach in Sec.~\ref{sec-theory-NEMD}.  The results are presented in
Sec.~\ref{sec-Alkane-Variation}. In particular, Sec.~\ref{sec-Alkane-Geometry}
deals with different absorption sites of the alkanedithiols on the electrodes
and different electrode orientations. Next, we investigate the influence of
gauche defects on transport in Sec.~\ref{sec-Alkane-Defects} and how they
respond to strain. Sec.~\ref{sec-Alkane-NEMD} adds an analysis of temperature
effects via the NEMD approach to the previous two DFT-NEGF studies. We close
with the conclusions in Sec.~\ref{sec-conclusions}.

\section{Theoretical methods}\label{sec-theory}

In order to give a comprehensive picture, we take advantage of two
complementary approaches to calculate the thermal conductance of
single-molecule junctions. To be precise, we first study the phase-coherent
phonon transport in the spirit of the Landau-B\"uttiker scattering theory
using NEGF techniques, where the device-specific parameters are obtained from
DFT as implemented by us and described in
Refs.~\cite{Buerkle2015,Kloeckner2019}. The phonon thermal conductance of the
nanoconstriction is calculated as a linear-response property based on the
harmonic approximation. Anharmonic scattering, for instance from phonon-phonon
interactions, is explicitly neglected, and the geometries are obtained as
energetically favorable configurations at temperature $T=0$.

These studies are complemented by NEMD simulations. In comparison to the
ab-initio approach, this method offers several advantages. NEMD naturally
considers the nonequilibrium situation and temperature gradients inside the
junction, i.e., the heat transport can be described for high applied
temperature differences between left and right electrodes. Since the
experiments measure the thermal conductance in the linear-response regime
\cite{Cui2019}, this is not relevant however, and we apply NEMD and DFT-NEGF
in that limit. Whereas the DFT-NEGF methodology is restricted to a few
hundreds of atoms for practical computational efforts, with the molecular
dynamics (MD) simulations
atom numbers on the order of many thousands can easily be
handled. Importantly, anharmonic effects are taken into account in a
non-perturbative manner. These anharmonic effects lead to phonon-phonon
interactions or the temperature-dependent lattice expansion. The
time-dependent simulations imply a sampling of the geometrical phase space.
The major drawback can be seen in the need for empirical interatomic
potentials, which are not universal and often optimized for special
purposes. This becomes most evident, when interface structures are considered
at the atomic scale. Due to lack of chemical information conventional
potentials then tend to fail to describe geometries reliably. Although there
are improved potentials, they come along with higher computational costs. In
contrast, DFT-based methods are known to provide reliable geometries at atomic
scales. Since MD uses classical statistics, another disadvantage is that
quantum effects are not taken into account. This limits the applicability to
thermal energies above the material-specific highest phonon energies. For
gold, our DFT calculations yield a cutoff at around 20~meV, which is on the
order of the Debye energy of 15~meV. This corresponds to temperatures above
around 200~K, justifying the application in this work.

In short, DFT-NEGF and NEMD approaches complement each other. Therefore both
will be used in our room-temperature study. In the following, we will present
them in greater detail.

\subsection{Phonon transport in the Landau-B\"uttiker scattering theory using DFT parameters}\label{sec-theory-DFT+NEGF}

For the calculation of heat transport via phonons in the phase-coherent
regime, we consider the Hamiltonian due to small displacements $\{Q_{\xi}\}$
of atoms away from their equilibrium position $\{R_{\xi}^{(0)}\}$
\begin{equation}
  \hat H = \frac{1}{2} \sum_{\xi} \hat{p}_{\xi}^{2} + \frac{1}{2\hbar^{2}} 
  \sum_{\xi\chi} \hat{q}_{\xi} K_{\xi\chi} \hat{q}_{\chi}.
\end{equation}
In this expression we have introduced the mass-weighted displacement
operators $\hat{q}_{\xi} = \sqrt{M_{\xi}}\hat{Q}_{\xi}$ and the mass-scaled
momentum operators $\hat{p}_{\xi}=\hat{P}_{\xi}/\sqrt{M_{\xi}}$.  They obey
the usual commutation relations: $[\hat{q}_{\xi},\hat{p}_{\chi}] =
\mbox{i}\hbar\delta_{\xi\chi}$ and $[\hat{q}_{\xi},\hat{q}_{\chi}] =
     [\hat{p}_{\xi},\hat{p}_{\chi}]=0$, where $\xi=(j,c)$ denotes a Cartesian
     component $c=x,y,z$ of atom $j$ at position $\vec{R}_{j} =
     \vec{R}_{j}^{(0)}+\vec{Q}_{j}$. As the system-specific quantity we
     calculate the dynamical matrix $K_{\xi\chi} = \hbar^{2}
     \partial_{\xi\chi}^{2}E_{\rm DFT}/\sqrt{M_{\xi}M_{\chi}}$ as the
     mass-weighted second derivative of the DFT total ground state energy
     $E_{\rm DFT}$ with respect to the Cartesian atomic coordinates.

For the transport calculation we divide the dynamical matrix into different
subsystems, namely a central (C) scattering region and the two semi-infinite
left (L) and right (R) electrodes
\begin{equation}
  \boldsymbol{K}=\left(\begin{array}{ccc}
    \boldsymbol{K}_{\mathrm{LL}} & \boldsymbol{K}_{\mathrm{LC}} & \boldsymbol{0}\\
    \boldsymbol{K}_{\mathrm{CL}} & \boldsymbol{K}_{\mathrm{CC}} & \boldsymbol{K}_{\mathrm{CR}}\\
    \boldsymbol{0} & \boldsymbol{K}_{\mathrm{RC}} & \boldsymbol{K}_{\mathrm{RR}}
  \end{array}\right). 
  \label{eq-Kmatrix}
\end{equation}
In our implementation we extract the elements
$\boldsymbol{K}_{X\text{C}},\boldsymbol{K}_{\text{C}X},\boldsymbol{K}_{\text{CC}}$
with $X=\text{L,R}$ from an extended central cluster, while the elements
$\boldsymbol{K}_{\text{LL}}$ are obtained from a separate bulk calculation. To
account for the acoustic sum rule $K_{jc,jc'}=-\sum_{j'\neq j}K_{jc,j'c'}
\sqrt{M_{j'}/M_j}$, where $j,j'$ indicate atoms and $c,c'$ Cartesian
components, we correct the surface elements as described in
Ref.~\cite{Kloeckner2019}.  Since we assume a vanishing coupling between the
two leads, the thermal conductance can be calculated as the linear-response
coefficient in a heat current formula \cite{Rego1998, Mingo2003, Yamamoto2006}
and reads
\begin{equation}
  \kappa_{\rm pn}(T) = \frac{1}{h} \int_{0}^{\infty} E \tau_{\rm pn}(E)
  \frac{\partial n(E,T)}{\partial T} \mathrm{d}E.
  \label{eq-kpn}
\end{equation}
Here $n(E,T)=[\exp(E/k_{\rm B}T)-1]^{-1}$ represents the Bose function and
$\tau_{\rm pn}(E)$ is the phonon transmission function. In the framework of
NEGF it is given by \cite{Mingo2003,Buerkle2015}
\begin{equation}
  \tau_{\rm pn}(E) = \mathrm{Tr} \left[ \boldsymbol{D}_{\mathrm{CC}}^{\textrm{r}}(E) 
    \boldsymbol{\Lambda}_{\textrm{L}}(E) \boldsymbol{D}_{\mathrm{CC}}^{\textrm{a}}(E)
    \boldsymbol{\Lambda}_{\textrm{R}}(E)\right],
  \label{eq-tauph}
\end{equation}
where $\boldsymbol{D}_{\mathrm{CC}}^{\textrm{r,a}}(E)$ are the retarded and
advanced phonon Green's functions of the center.  They are defined as
$\boldsymbol{D}_{\rm CC}^{\rm a}(E) = \boldsymbol{D}_{\rm CC}^{\rm
  r}(E)^{\dagger}$ and
\begin{equation}
  \boldsymbol{D}_{\mathrm{CC}}^{\mathrm{r}}(E) = \left[\left(E+i\eta
    \right)^{2}\boldsymbol{1}_{\textrm{CC}} - \boldsymbol{K}_{\textrm{CC}} -
    \boldsymbol{\Pi}_{\textrm{L}}^{\mathrm{r}}(E) -
    \boldsymbol{\Pi}_{\textrm{R}}^{\mathrm{r}}(E) \right]^{-1} 
\end{equation}
with the infinitesimal parameter $\eta>0$. The scattering rate matrices
\begin{equation}
  \boldsymbol{\Lambda}_{X}(E)= i
  \left[\boldsymbol{\Pi}_{X}^{\mathrm{r}}(E)-\boldsymbol{\Pi}_{X}^{\mathrm{a}}(E)\right]
\end{equation}
are related to the corresponding embedding self-energies
\begin{equation}
  \boldsymbol{\Pi}_{X}^{\mathrm{r}}(E) = \boldsymbol{K}_{\textrm{C}X} 
  \boldsymbol{d}_{XX}^{\mathrm{r}}(E) \boldsymbol{K}_{X\mathrm{C}},
\end{equation}
which describe the coupling between the C region and electrode $X$. In the
expressions $\boldsymbol{\Pi}_{X}^{\textrm{a}}(E) =
\boldsymbol{\Pi}_{X}^{\textrm{r}}(E)^{\dagger}$ and $\boldsymbol{d}_{XX}^{\rm
  r}(E)= [ (E+i\eta)^{2}\boldsymbol{1}_{XX} -
  \boldsymbol{K}_{\textrm{XX}}]^{-1}$ is the surface Green's function of lead
$X=\text{L},\text{R}$, which we calculate using the decimation technique
explained in Refs.~\cite{Pauly2008,Guinea1983}.

To better understand heat transport, we consider the decomposition of the
total phonon transmission of Eq.~(\ref{eq-tauph}) into contributions of
transmission eigenchannels, $\tau_{\rm pn}(E) = \sum_i \tau_{{\rm
    pn},i}(E)$. The coefficients $\tau_{{\rm pn},i}(E)$ are the eigenvalues of
the phonon transmission probability matrix $\boldsymbol t_{\rm pn}(E)
\boldsymbol t^{\dagger}_{\rm pn}(E)$, where $\boldsymbol t_{\rm pn}(E) =
\boldsymbol \Lambda^{1/2}_{\rm R}(E) \boldsymbol D_{\rm CC}^{\rm r}(E)
\boldsymbol \Lambda^{1/2}_{\rm L}(E)$ is the phonon transmission amplitude
matrix.  In this way on gets insight into the number of eigenchannels, which
are important for heat transport at a specific energy, and a visualization of
the transmission eigenchannels yields information on how the vibrational
energy is transported through the nanojunctions. For a detailed discussion of
the methodology, we refer to Ref.~\cite{Kloeckner2018}.

Another useful quantity is the cumulative thermal conductance
\begin{equation}
  \kappa^{\mathrm{ c}}_{\mathrm{ pn}}(E,T)= \frac{1}{h} \int_{0}^{E} E' \tau_{\rm ph}(E') 
  \frac{\partial n(E',T)}{\partial T} \mathrm{d}E',
  \label{eq-kcpn}
\end{equation}
defined as the thermal conductance at temperature $T$ due to phonon modes up
to the given energy $E$. It provides information on how vibrational modes
contribute to the total thermal conductance.

For the DFT calculations we use TURBOMOLE V7.1
\cite{TURBOMOLE,Deglmann2002,Deglmann2004}, employ the Perdew-Burke-Ernzerhof
PBE exchange-correlation functional \cite{Perdew1992,Perdew1996} and the
“default2” basis set of split-valence-plus-polarization quality def2-SV(P)
\cite{Weigend2005} together with the corresponding Coulomb fitting basis
\cite{Weigend2006}. In order to accurately determine the vibrational energies
and force constants, we use very strict convergence criteria. In particular,
total energies are converged to a precision of better than $10^{-9}$~a.u.,
whereas geometry optimizations are performed until the change of the maximum
norm of the Cartesian gradient is below $10^{-5}$~a.u.. For the transmission
calculations we use $32\times32$ $k$-points transverse to the direction of
transport and $\eta=10^{-5}$~a.u..

\subsection{Phonon transport in NEMD}\label{sec-theory-NEMD}

\begin{figure}
  \centering \includegraphics[width=1.0\columnwidth]{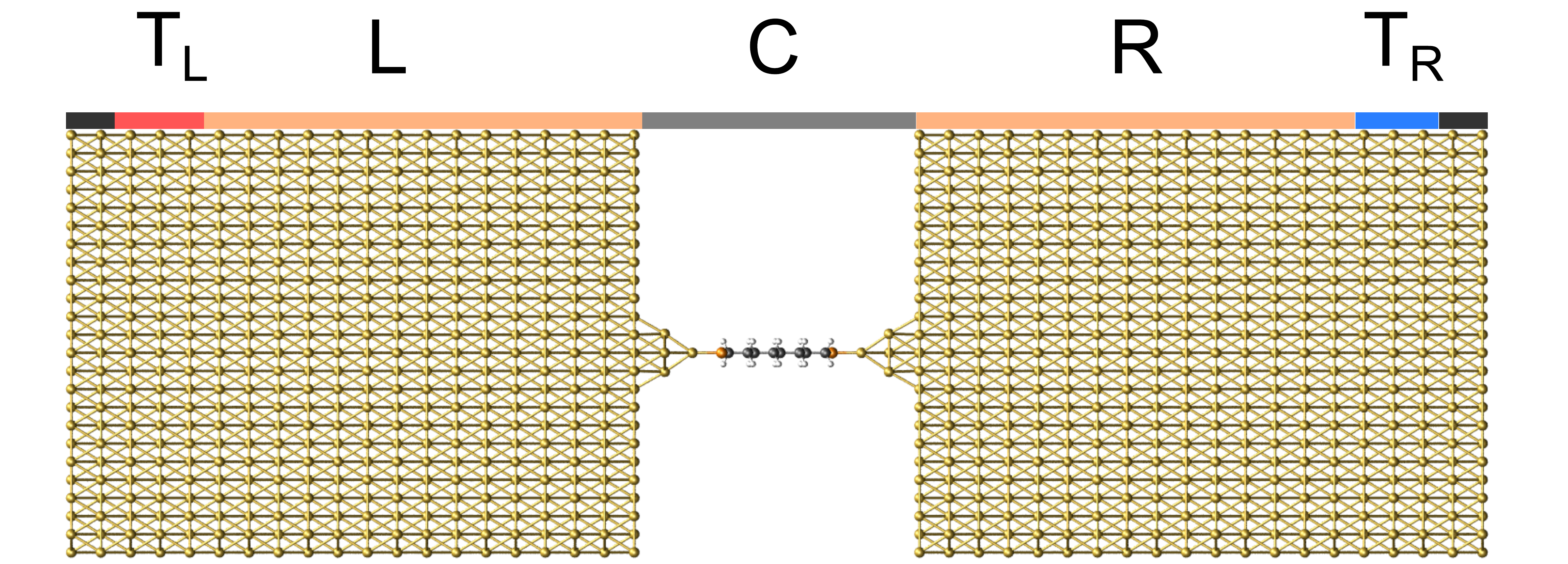}
  \caption{Geometry used in the NEMD simulations. The outermost atoms (marked
    by a black line above the corresponding atomic planes) are fixed to
    prevent a collapse of the system. They are followed by the atoms that are
    connected to the thermostats (red and blue) and held at constant
    temperatures of $T_{\text{L}}$ and $T_{\text{R}}$. The flexible electrode
    reservoir atoms (orange) on the left (L) and right (R) are bridged by a
    central (C) junction part (gray), containing gold tip structures and the
    alkane molecule.}
  \label{fig-NEMD-Setup}
\end{figure}

As an alternative to the calculation of the thermal conductance based on the
DFT-NEGF methodology, MD offers different possibilities to predict the heat
current at the atomistic level. Generally speaking, the MD methods can be
classified into two different subgroups, namely equilibrium MD and NEMD
\cite{Schelling2002}. While in equilibrium MD one computes transport
properties from correlation functions using the Green-Kubo formalism, in NEMD
the system is driven out of equilibrium and the response is calculated
directly. In principle, with either method both
bulk~\cite{Schelling2002,Mueller-Plathe1997,Ciccotti1980} and
interface~\cite{Landry2009,Chalopin2012,Barrat2013,Merabia2012} thermal
conductivities can be calculated. However, all of the MD methods require a
certain minimal amount of atoms to avoid finite-size effects
\cite{Sellan2010,Liang2014,Wang2017}.

In this work we use a direct NEMD approach, where the nonequilibrium situation
is realized by maintaining two spatially separated regions at a constant
temperature difference using thermostats, see Fig.~\ref{fig-NEMD-Setup}. Our
simulations are done with the LAMMPS software code \cite{Plimpton1995}.

A crucial factor in MD simulations is the choice of the interatomic
interaction potentials. Calculations of heat transfer through hybrid
metal-organic interfaces have used a Morse potential to describe the
sulfur-gold bond \cite{Luo2010, Soussi2015}. This kind of parametrization is
useful to characterize the binding situation in a self-assembled
monolayer~\cite{Luo2010, Ong2014, Majumdar2015}. In single-molecule junctions,
however, we find that these potentials are incapable of describing the
chemical bond at the metal-molecule contact correctly but lead to an
overcoordination at the sulfur atom, which tends to bind to too many gold
atoms. Additionally, covalent intramolecular interactions are often treated in
the harmonic approximation \cite{Luo2010,Soussi2015,Ong2014,Majumdar2015}. In
this work, we employ a reactive force field (REAXFF)
\cite{Russo2011,Senftle2016}, which is based on the concept of bond order to
model bonded and non-bonded interactions between atoms. Importantly, it can
describe both bond breaking and bond formation, and all interatomic
interactions in the systems are anharmonic.

Figure~\ref{fig-NEMD-Setup} shows a starting geometry of our NEMD
simulations. We employ periodic boundary conditions perpendicular to the
direction of transport and fixed boundary conditions parallel to it. The
outermost atoms on each side, indicated by the black line above the geometry,
are held fixed to avoid a collapse of the system. These regions are followed
by several atom rows, marked by the red and blue lines, where thermostats are
applied that keep these regions at constant temperatures. In detail we use two
velocity rescale thermostats (with \emph{fraction} factor $\alpha=0.8$ und
\emph{window}$= 0.01$~K) to maintain temperatures at $T_{\text{L}}$ and
$T_{\text{R}}$, as described in Ref.~\cite{Moehrle2019}. The largest part of
the simulation box consists of the L and R electrode atoms, indicated by the
orange lines. To avoid finite-size effects, we have included in these regions
so many atoms, that the phononic density of states on each side closely
resembles that of bulk.  Similar to the DFT-NEGF formalism we define all atoms
in between the leads as C part, marked in gray.

Since we are investigating a nonequilibrium situation, we need to take
thermal expansion into account. The temperature difference in the L and R
regions leads to a corresponding difference in lattice constants in the
electrodes. In order to distribute strain effects equally over the whole
junction structure, we therefore determine the volume of the simulation box
from the effective lattice constant
\begin{equation}
  a_{\mathrm{eff}}=\sqrt[3]{\dfrac{a_{\mathrm{L}}^3+a_{\mathrm{R}}^3}{2}},
\end{equation}
where $a_X$ are bulk lattice constants at temperature $T_X$. 

In order to calculate a thermal conductance that is comparable to the DFT-NEGF
method, the starting point for the C part in the NEMD simulations coincides
with the DFT geometry of junction type 1 (JT1), compare
Figs.~\ref{fig-NEMD-Setup} and \ref{fig-ContactGeo}. Large electrodes,
oriented along the (111) direction, in the L and R regions are matched to the
inner gold pyramid structures. While this procedure ensures geometries in C
that are similar to JT1, the junction is likely to be in a different strain
situation. This is attributed to differences in equilibrium interatomic
distances between DFT and the REAXFF. Additionally, the thermal expansion
modifies interatomic equilibrium distances in the NEMD simulations, as
discussed in the previous paragraph. The L and R regions in
Fig.~\ref{fig-NEMD-Setup} are five unit cells long, when seen along the
transport direction. Regions, where thermostats are applied, are one unit cell
thick, and the simulation box is enclosed by two rows of fixed atoms on each
side. In the transverse directions we use 12 unit cells, oriented along
$(\bar{1}\bar{1}2)$ and $(1\bar{1}0)$ crystallographic directions,
respectively. The NEMD junction geometry consists on each side of 384 fixed
atoms and 576 atoms in the parts coupled to thermostats. Furthermore, 2880
atoms reside in L and R parts, respectively. With the 4 gold atoms in the
pyramidal gold structures of each side and the variable number of atoms in the
dithiolated alkane chains C4-C10, this adds up to more than 7700 atoms.

For the integration of Newton's equation of motion, we use the velocity-Verlet
algorithm \cite{Swope1982}. We perform the NEMD simulations as follows. Apart
from the fixed atom regions, random velocities are assigned to all atoms in
the junction to yield a Gaussian distribution centered at 300~K. We start with
1~ns of equilibration with temperatures in the thermostats set to
$T_{\text{L}}=T_{\text{R}}=300$~K. Next we increase the temperature of the
left thermostat $T_{\text{L}}$ by the desired temperature difference $\Delta
T$ to $T_{\text{L}}=T_{\text{R}}+\Delta T$ over 1~ns. After an additional
equilibration time of 1~ns, we perform a time evolution over 4~ns, which
constitutes the actual simulation run. The time step chosen for the C4-derived
junction is 0.5~fs and 1~fs for C6-C10. In the NEMD simulations the separation
between the fixed layers is held constant, i.e.\ no pulling process is
performed.

The thermal conductance is calculated according to
\begin{equation}
  \kappa_{\mathrm{pn}}= \frac{J}{\Delta T}=\frac{\Delta E}{\Delta t \Delta T},\label{eq-kpn-NEMD}
\end{equation}     
where $J=\Delta E/\Delta t$ is the steady-state heat current. Overall,
$\kappa_{\text{pn}}$ is thus obtained from the energy $\Delta E=(-\Delta
E_{\text{L}}+\Delta E_{\text{R}})/2$ supplied to the system by the thermostats
L and R per time step $\Delta t$, divided by the temperature difference
$\Delta T=T_{\text{L}}-T_{\text{R}}$. Details of the procedure are described
in Ref.~\cite{Moehrle2019}. In practice, we cumulate the output of the
thermostats $\Delta E$ over the simulation time $t=N\times\Delta t$ of $N$ MD
time steps and extract the heat current $J$ as the slope of a linear fit to
$\Delta E^{\text{c}}(t)=\sum_{i=1}^N\Delta E(i \times \Delta t)$. In the
simulations, we print out $\Delta E^{\text{c}}(t)$ every 2~ps.

\section{Variability of the phonon thermal conductance}\label{sec-Alkane-Variation}

Let us in the following discuss the phonon thermal conductance of
single-molecule junctions formed from dithiolated alkanes with 4-10 CH$_2$
units, hereafter named C4-C10, contacted by two gold electrodes. Since the
electronic contribution is negligible, the phonon thermal conductance is the
relevant property that is measured in novel heat transport experiments
\cite{Cui2019}. We quantify here the influence that different geometrical
variations exert on the phonon thermal conductance of single-molecule
junctions. These are the adsorption sites of the sulfur anchors on the gold
electrodes and the electrode orientation in Sec.~\ref{sec-Alkane-Geometry},
and the influence of gauche defects and strain in
Sec.~\ref{sec-Alkane-Defects}. In these two sections, we use the DFT-NEGF
approach. Finally, in Sec.~\ref{sec-Alkane-NEMD}, we study the effect of
thermally induced geometry variations with the help of the NEMD method.

\subsection{Adsorption site and electrode orientation}\label{sec-Alkane-Geometry}

\begin{figure}[t]
  \begin{center} \includegraphics[width=1.0\columnwidth]{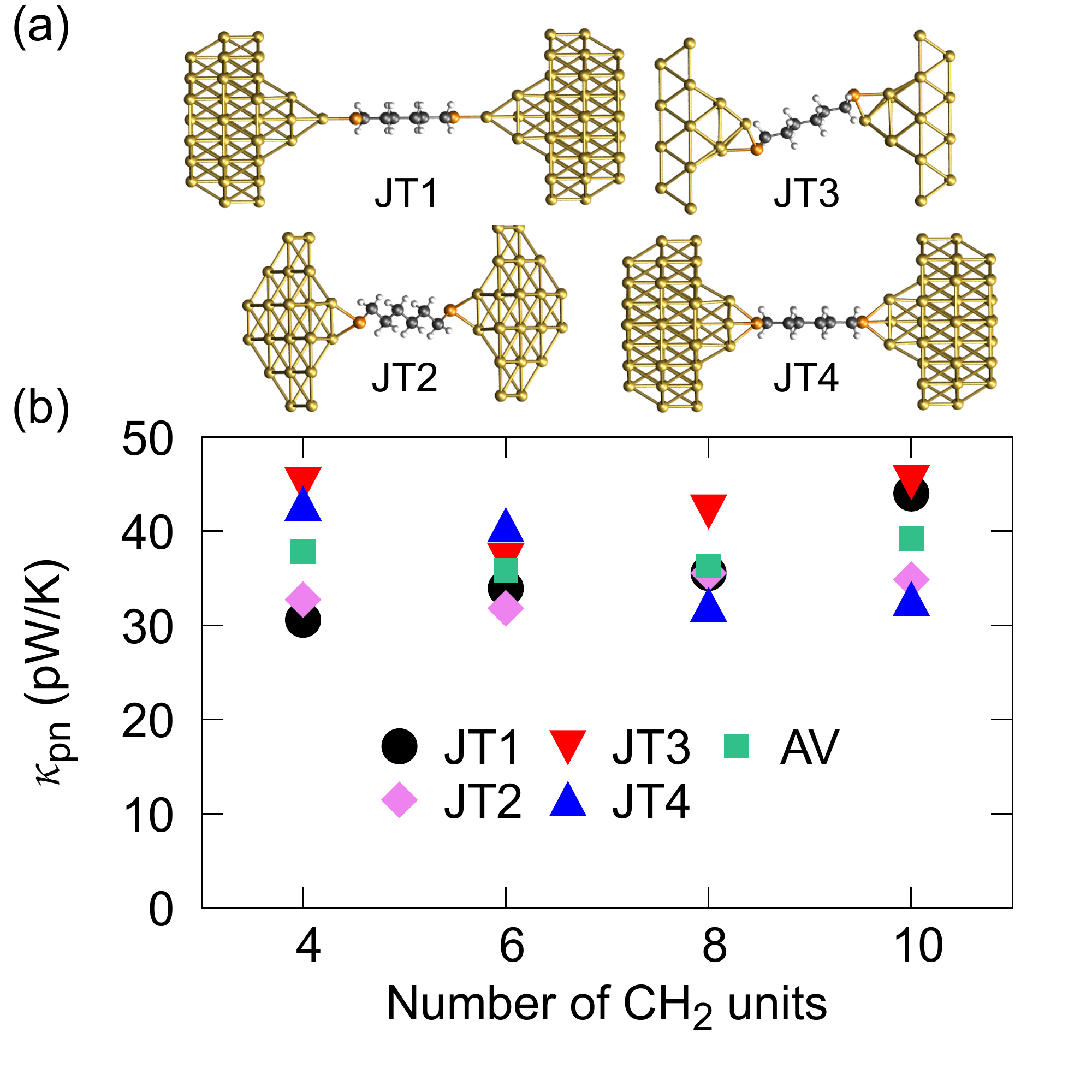} \end{center}
  \caption{(a) Different types of junction geometries (JT1-JT4) used for
    estimating the variability of $\kappa_{\text{pn}}$ in dependence on the
    molecule-electrode contact and the electrode orientation, here shown for
    C6. (b) Phonon thermal conductance for each junction type, as determined
    with the DFT-NEGF method, and the average taken over the values of
    JT1-JT4.}
  \label{fig-ContactGeo}
\end{figure}

To better understand the size of variability of the phonon thermal conductance
in single-molecule junctions, we start by determining $\kappa_{\text{pn}}$ for
different junction types with the DFT-NEGF approach. The junctions have been
constructed by varying the binding position of the molecular sulfur termini on
the gold electrodes as well as the orientation of the metal electrodes with
respect to the heat transport direction. Sulfur atoms on both ends are bonded
to a single gold tip atom in atop position for JT1, two gold atoms in bridge
position for JT2 and JT3, and three gold atoms in a hollow position in
JT4. The electrode orientations are $(111)$ for JT1 and JT4, $(110)$ for JT2
and $(100)$ for JT3.  The various contact geometries are displayed in
Fig.~\ref{fig-ContactGeo}(a) for C6. To provide a meaningful comparison among
different junction types, JT1-JT4 use fully extended, straight alkane
chains. They are built based on the same protocol, which minimizes strain
effects by first optimizing the molecule connected to one gold tip structure
and then attaching a gold tip on the other side symmetrically and reoptimizing
the geometry of the full molecular junction. Within one junction type, all the
alkanes of different length show a comparable geometry. The computed phononic
thermal conductance is depicted in Fig.~\ref{fig-ContactGeo}(b) and found to
be in the range of 30 to 50~pW/K with mean values around 35 to 40~pW/K. We
note that this data has been presented in ``Extended Data Table 1'' of
Ref.~\cite{Cui2019}.  As discussed there, the thermal conductances are
somewhat larger than experimentally detected, since we assume a situation with
low mechanical strain. Instead, at the experimentally relevant breaking point,
the mechanical tension can pull out gold atoms from the soft electrodes,
reducing $\kappa_{\text{pn}}$.

\begin{figure*}[t]
  \begin{center} \includegraphics[width=1.0\textwidth]{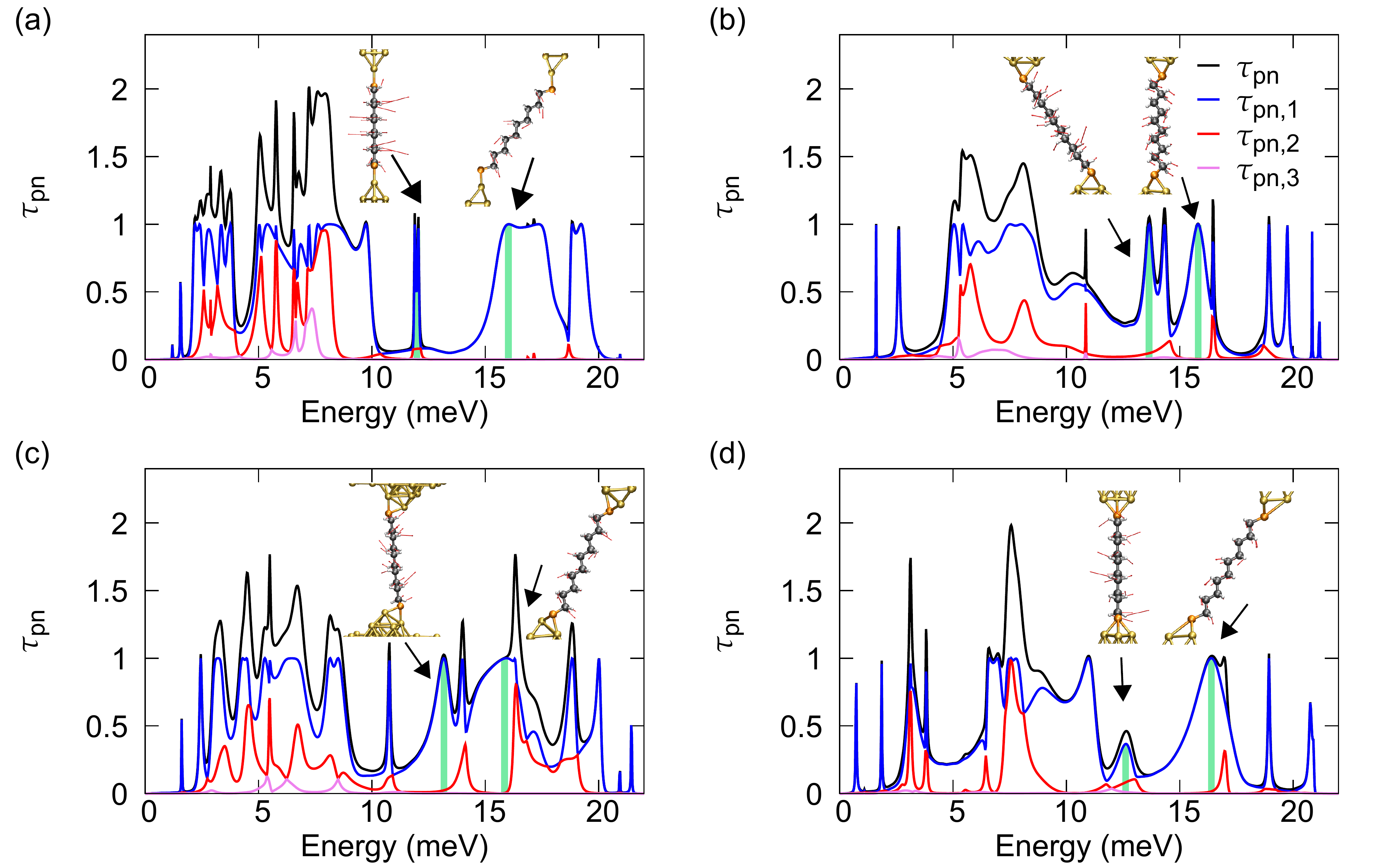} \end{center}
  \caption{Total phonon transmission $\tau_{\text{pn}}(E)$ resolved into
    eigenchannel contributions. For JT1-JT4 the transmissions
    $\tau_{\text{pn},i}(E)$ of the three most transmissive eigenchannels
    $i=1,2,3$ are depicted as a function of energy for C10 in panels (a)-(d),
    respectively. In the inset of each panel, a static picture of the
    eigenchannel $i=1$ with the highest transmission is shown for two energies
    that are indicated by a vertical green line.}
  \label{fig-TransGeo}
\end{figure*}

The length-independent thermal conductance can be understood by studying the
phonon transmission \cite{Kloeckner2016,Cui2019}, which gives the probability
of a phonon to be transmitted elastically, i.e.\ at a specified energy, from
one electrode to the other through the alkane molecule. As visible in
Fig.~\ref{fig-TransGeo} for C10, $\tau_{\text{pn}}(E)$ exhibits resonances
with positions and widths that vary with contact geometry. The transmission
vanishes for energies above around $E_{\text{max}}=20$~meV, which corresponds
to the highest phonon energies of gold and is comparable to the Debye energy
of 15~meV \cite{Ashcroft1976}. Through the nearly constant factor $\partial
n(E,T)/\partial T$ in Eq.~(\ref{eq-kpn}), the transmissions in the range from
0 to $E_{\text{max}}$ contribute almost equally to $\kappa_{\text{pn}}$ at
$T=300$~K. Transmission resonances in this energy interval arise from
center-of-mass motions of the alkanes between the gold electrodes or molecular
vibrations with low energy. The longer the molecule, the more low-energy
vibrational modes will overlap with the phonon density of states of gold,
leading to an increasing number of transmission resonances between 0 and
$E_{\text{max}}$. At the same time transmission resonances become more narrow
for the longer alkanes due to a reduced electrode-molecule linewidth
broadening. The balance between both effects provides an intuitive explanation
for the rather length-independent phonon thermal conductance, as computed here
and measured in Ref.~\cite{Cui2019}.

To provide more insight into the origin of the thermal conductance variations,
we resolve in Fig.~\ref{fig-TransGeo} the transmission through C10 for the
whole series of junction types into contributions of the three most
transmissive eigenchannels. Irrespective of the junction type we observe that
two eigenchannels dominate the transmission, while a third one gives only
small contributions. With respect to the position and width of the individual
transmission resonances, there is a lot of change between different
junctions. Since transmission resonances are linked to the vibrational
structure of the junctions, they will sensitively depend on the precise
contact geometry, in particular the coupling of the molecule to the gold
electrodes. As a general feature, we find however that molecular
center-of-mass motions tend to yield peaks in the lower part of the energy
range between 0 and $E_{\text{max}}$, while molecule-internal vibrations are
responsible for resonances in the upper part.

To quantify geometry-dependent changes of vibrational modes, we search for
eigenchannels with similar shape in the energy range of nonvanishing
transmission of Fig.~\ref{fig-TransGeo}. We restrict this analysis to the most
transmissive first eigenchannel of JT1 at energies of 12.05 and 16.02 meV. On
the molecule the vibrations involved can be classified as out-of-plane and
in-plane modes, respectively, if we consider the plane spanned by the
molecular backbone of sulfur and carbon atoms. Both eigenchannels show two
characteristic sign changes when going from one end of the molecule to the
other, which we use to identify the corresponding eigenchannels in the other
junctions types. As indicated in each panel by insets and vertical green
lines, we find the position of the corresponding modes at 13.65 and 15.80~meV
for JT2, 13.16 and 16.05~meV for JT3, and 12.78 and 16.43~meV for JT4. While
we observe only moderate shifts in energy, the peak widths as well as the peak
heights may differ substantially, which exemplifies the crucial impact of
contact details on phonon transport.

\subsection{Molecule-internal defects and mechanical strain}\label{sec-Alkane-Defects}

\begin{figure*}[bt]
  \includegraphics[width=0.9\textwidth,clip]{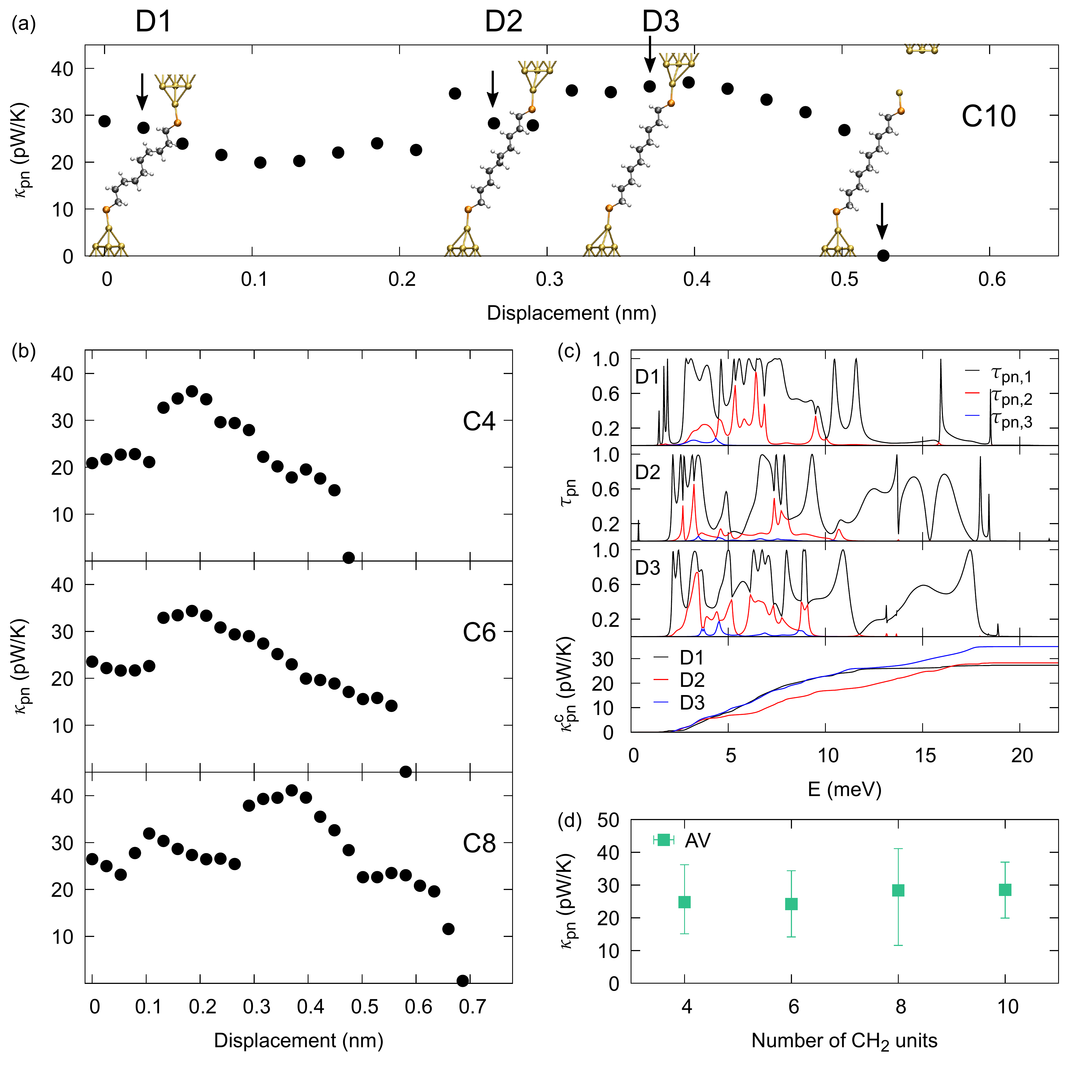}
  \caption{(a) Phonon thermal conductance $\kappa_{\text{pn}}$ vs.\ electrode
    displacement for the C10 chain containing gauche defects. Arrows indicate
    the thermal conductance of the four geometries visualized in the
    inset. (b) Corresponding pulling curves for the alkanes C4-C8. (c)
    Transmission resolved into eigenchannels for the three geometries labeled
    D1, D2, and D3 in panel (a), together with the cumulative thermal
    conductance of these junctions. (d) Thermal conductance of C4-C10,
    calculated as an average over the whole pulling curves shown in panels (a)
    and (b).}
  \label{fig-Defects}
\end{figure*}

In the previous section we focused on the thermal conductance of alkane chains
in a straight configuration. However, alkane chains are known to exhibit
geometrical defects, which can be thermally activated. In a study of
gold-alkane-gold junctions using ab-initio MD simulations \cite{Paulsson2009},
the authors related a peak in the histogram of the electrical conductance to
these defects. Furthermore they pointed out that upon stretching of the system
all defects vanish. For this reason the alkanes are expected to exhibit a
straight configuration before breaking of the contact. More recently these
defects have been shown to reduce the thermal conductance in comparison to
straight chains in a work employing graphene leads \cite{Li2015}. The authors
suggested to exploit the mechanism to realize a mechanically driven thermal
switch.

To test the influence of molecule-internal disorder on the phonon thermal
conductance of C4-C10, we generated chains containing defects by arbitrarily
varying torsional angles in the carbon backbone. As described in
Sec.~\ref{sec-Alkane-Geometry}, we set up the junction in the usual way by
first relaxing the twisted chain on top of one electrode and then
symmetrically attaching a gold pyramid by point mirroring at the free thiol
group. The geometries, which exhibit atomically sharp tips similar to JT1 in
Fig.~\ref{fig-TransGeo}, are then stretched to explore their response to
different strain conditions. We displace electrodes along the distance vector
connecting the gold tip atoms with increments of $0.5~\text{a.u.}\approx
0.265$~\AA.

Figure~\ref{fig-Defects}(a) shows the pulling of the alkane chain with 10
CH$_2$ units. Initially, the alkane exhibits four gauche defects. Since the
gauche defects appear in pairs, the alkane can be described as consisting of
three straight segments. In this initial stage the thermal conductance varies
smoothly around 20-30~pW/K, while the alkane adjusts continuously to the
increasing electrode separation. The region ends at a displacement of 0.24~nm
with a jump in $\kappa_{\text{pn}}$. Here the geometry of the alkane chain
changes, two gauche defects vanish, and the remaining pair splits the chain
into two straight segments. After two additional displacement steps, leading
to a decreased thermal conductance due to the induced strain, an additional
jump occurs at a displacement of 0.32~nm. From that moment on the chain is
defect-free. Further pulling leads basically to a continuous decrease of the
thermal conductance, before the contact finally breaks at an Au-Au bond. A
similar behavior is observed for the other chain lengths with 4-8 CH$_2$
segments, depicted in Fig.~\ref{fig-Defects}(b). Again defects in C4-C8 vanish
upon pulling, as it is evident from the discontinuous behavior of the thermal
conductance. In all examples the thermal conductance of the fully extended
configuration is larger than that of geometries with defects. We note that in
some cases studied in Fig.~\ref{fig-Defects}, Au chains form during the
pulling process, namely for the C4, C6 and C8 junctions. The formation of
chains depends sensitively on the starting configuration in the DFT
simulations, as discussed in Ref.~\cite{Cui2019}.

To better understand the origin of the reduction of $\kappa_{\text{pn}}$
through the torsional gauche defects, we show in Fig.~\ref{fig-Defects}(c) the
eigenchannel-resolved transmission of three examples with 4, 2, and 0
defects. The geometries of the junctions are depicted as insets, and their
thermal conductance values are indicated by arrows and labeled by increasing
displacement as D1, D2 and D3 in Fig.~\ref{fig-Defects}(a). In the two
examples with defects, D1 and D2, phonons within a different energy range are
responsible for the reduced conductance as compared to D3. In D1 this energy
range is located at about 13-17 meV, for D2 the reduced transmission results
from around 4-7~meV. This is particularly evident by analyzing the cumulative
thermal conductance. Here, the curves D1 and D3 almost coincide up to an
energy of around 13 meV but deviate above. A comparison of D2 and D3, on the
other hand, reveals the main difference to start occurring around 4~meV,
whereas after 7~meV the offset in $\kappa_{\text{pn}}^{\text{c}}$ stays more
or less constant.

Finally, we show in Fig.~\ref{fig-Defects}(d) the mean thermal conductance of
C4-C10, calculated by averaging over all the thermal conductance values in the
contact regime of the pulling process with $\kappa_{\text{pn}}>
0.1$~pW/K. Highest and lowest thermal conductance values in this range are
indicated through arrow bars. The mean thermal conductance represents an
average over several strain conditions and geometrical configurations, and it
shows a nearly length-independent value of 20-30 pW/K. Of course, the starting
point for taking the average is somewhat arbitrary in this case, since it is
determined by the chosen starting configuration. Interestingly, however, the
values predicted for the thermal conductance are similar to what we have
computed in Ref.~\cite{Cui2019} at the point of contact rupture. Indeed,
Fig.~\ref{fig-Defects}(a) and \ref{fig-Defects}(b) show that junctions with
defects in the alkane chain can feature a similar thermal conductance as
junctions with straight chains before the point of rupture.

\subsection{Thermally induced geometry changes}\label{sec-Alkane-NEMD}

\begin{figure}[tb]
  \begin{center} \includegraphics[width=1.0\columnwidth,clip]{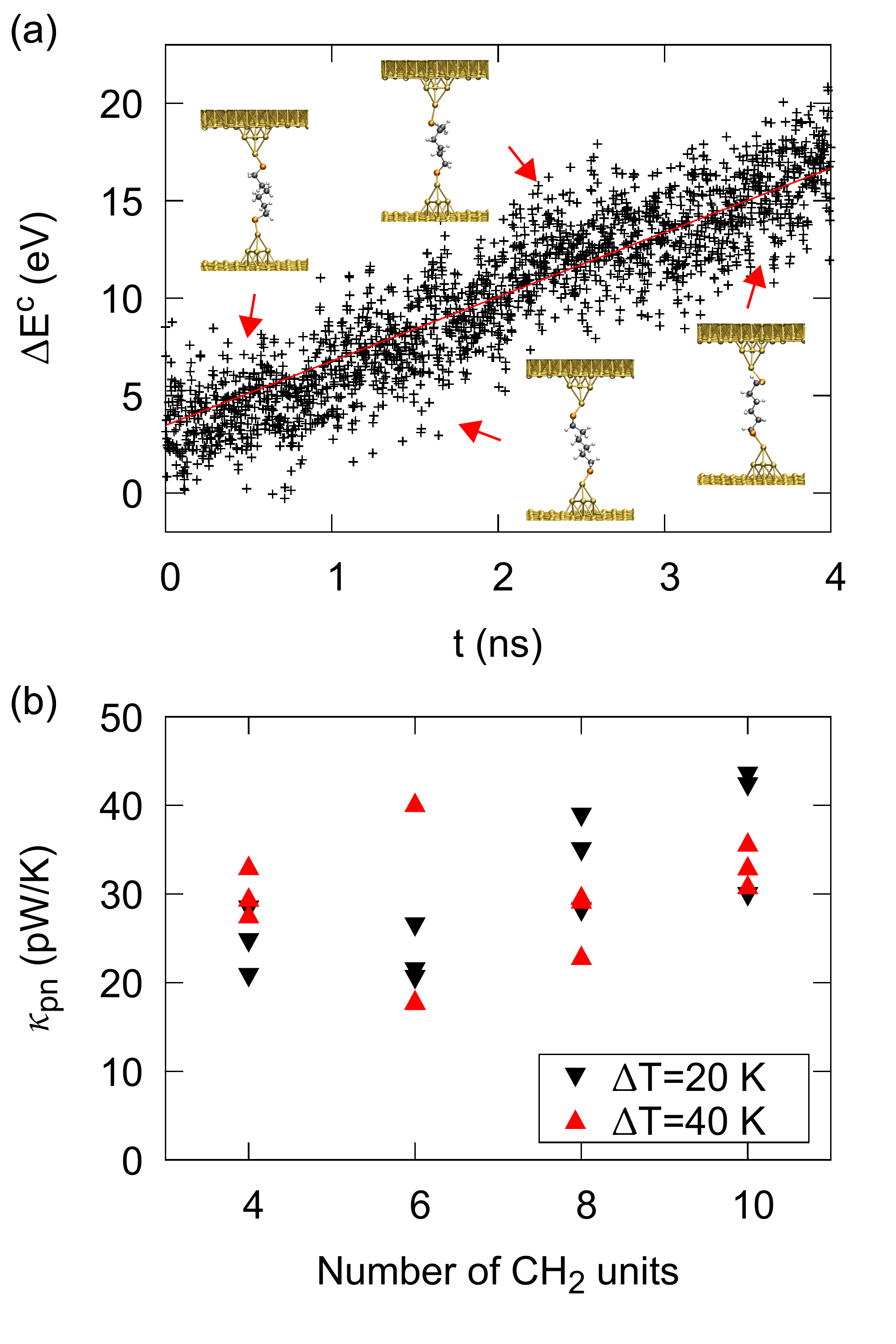} \end{center}
  \caption{(a) Cumulated output of the thermostats $\Delta E^{\text{c}}$ over
    time for a C6 junction. The red line shows the linear fit used to
    calculate the thermal current $J$. For each nanosecond interval a
    representative geometry is depicted. The non-zero value of $\Delta
    E^{\text{c}}$ at $t=0$ is due to the previous equilibration and
    temperature rap. (b) Calculated thermal conductance for C4-C10 using
    temperature gradients of $\Delta T=20,40$~K. For each molecule and
    temperature gradient, data from three different MD runs is displayed. For
    C6 at $\Delta T=40$~K two points fall together onto the the low thermal
    conductance value of around 18~pW/K.}
  \label{fig-AlkaneNEMD}
\end{figure}

To close the analysis, we study temperature-induced geometry changes of
single-molecule junctions and their effect on the thermal conductance using
NEMD simulations. The chosen interatomic interaction potentials are a crucial
point and need to describe reliably the molecular geometry, those of the
metallic electrodes and at the same time the metal-molecule interface. Only in
this way they can reproduce the vibrational modes that determine the phonon
heat transport. As discussed in Sec.~\ref{sec-theory-NEMD}, we construct
junction geometries for the NEMD simulations, see Fig.~\ref{fig-NEMD-Setup},
which show a similar C part as JT1 of Fig.~\ref{fig-ContactGeo}, studied with
the DFT-NEGF approach. We do so to ensure comparability between both
methods. We use the REAXFF of Ref.~\cite{Jaervi2011, Bae2013,Senftle2016} in
our study. It has been optimized for Au-S-C-H systems \cite{Bae2013}, yields
anharmonic interatomic interactions in the whole system, and is able to
describe chemical bond formation and dissociation. Overall, we find it to
yield reasonable junction geometries.

In Fig.~\ref{fig-AlkaneNEMD}(a) the cumulated output of the thermostats is
displayed as a function of simulation time for a C6 junction. As explained in
Sec.~\ref{sec-theory-NEMD}, we calculate the thermal conductance as the slope
of the linear fit, indicated by the red line in Fig.~\ref{fig-AlkaneNEMD}(a),
divided by the applied temperature difference $\Delta T$, see
Eq.~(\ref{eq-kpn-NEMD}). It is visible that the output of the thermostats show
fluctuations on the order of roughly $\pm 5$~eV. Let us note that $\Delta
E^{\text{c}}$ is an extensive quantity that is proportional to the number of
atoms in the regions coupled to the thermostats, and it thus amounts only to
$\pm 8.7$~meV/atom. For each interval of a nanosecond simulation time, we show
a sample geometry as an inset. Interestingly, apart from the extended chain at
the start the alkanes can exhibit different geometrical defects. Temperature
fluctuations in the large atom reservoirs in the L and R regions in
Fig.~\ref{fig-NEMD-Setup} lead to an effective modulation of the electrode
separation in the C part, which can stretch or squeeze the molecule. The
thermal conductance at ambient temperatures arises thus as an average over
geometrical configurations, including chains with and without
defects. Although shown here only for C6, this holds also for the other
molecules.

The thermal conductance for two different temperature differences $\Delta
T=20,40$~K is plotted in Fig.~\ref{fig-AlkaneNEMD}(b) for C4-C10. First of all
we note that there is no significant difference related to the applied
temperature gradient, meaning that we are in the linear-response
regime. Overall the conductances range from 20-40~pW/K. This is somewhat
smaller than the thermal conductance determined for the straight alkane chains
in the different geometries studied in Fig.~\ref{fig-ContactGeo}. In
particular, it is lower than the conductances of JT1, from which the NEMD
geometries are derived and which we predict to lie between 30 and
45~pW/K. Conversely, the results are in good agreement with the values of
$\kappa_{\text{pn}}$, sampled in the pulling curves that started from
defective chains in Sec.~\ref{sec-Alkane-Defects}. For C10 we observe that due
to the mentioned mismatch of interatomic distances between DFT and the REAXFF
at finite temperature, the geometry at one side changes from an atomically
sharp tip to a blunt shape. This corresponds to a higher stress configuration
of the alkane chain, or, in analogy with the DFT studies, to a lower
probability of defect formation. Indeed, $\kappa_{\text{pn}}$ for C10 is
larger than for the other molecules in the NEMD study and in closer agreement
with the DFT results of the straight chain.

Overall, we find a good agreement between DFT-NEGF and NEMD methods for the
thermal conductance. While the results of both approached will depend on the
chosen exchange correlation potential and the selected force field, the
consistency indicates that in these single-molecule nanojunctions, anharmonic
effects are weak. For this reason, vibrations in the C part behave like waves
that proceed elastically through the system at a fixed energy. Since
relaxation of energy and dephasing will happen inside the metal reservoirs,
the phonon heat transport over the nanoconstriction can also be characterized
as phase-coherent.

\section{Conclusions}\label{sec-conclusions}
In this paper, we presented a comprehensive overview of variations in the
phonon thermal conductance of alkane single-molecule junctions at room
temperature. For this purpose, we used different methods: A combination of DFT
and NEGF to study the thermal conductance in the phase-coherent, harmonic
regime and an approach based on classical NEMD to include the effects of
temperature and anharmonicity. Since we have shown in Ref.~\cite{Cui2019} that
the electronic contribution to the thermal conductance plays no role for C4
and longer alkanes, the phonon thermal conductance is equivalent to the full
thermal conductance apart from radiative effects. Due to the particular
measurement scheme, established in Ref.~\cite{Cui2019}, where the molecular
junctions spontaneously break at a fixed electrode separation, radiative
contributions cancel out and the phonon thermal conductance can thus be
uniquely identified.

As the first point, we estimated the variations of $\kappa_{\text{pn}}$ due to
different binding positions of sulfur anchors to the gold electrodes,
including atop, bridge or hollow positions, as well as due to the orientation
of the leads, examining alignments of $(111)$, $(110)$ and $(100)$ crystal
directions with the transport direction. These simulations were done with
extended alkane chains and the DFT-NEGF method. We found the average thermal
conductance to be independent of length within a range of 35-40 pW/K with
variations on the order of $\pm 5$~pW/K. By studying the eigenchannel-resolved
phonon transmission, we could identify peaks originating from similar
molecular vibrations. This served to demonstrate the sensitivity of individual
modes to the precise contact geometry. Next, we looked at the influence of
molecule-internal disorder. For this purpose, we generated a set of junction
geometries, where the alkane contains at least one gauche
defect. Subsequently, we varied the distance between the electrodes to
investigate the response of the defects to different strain conditions and the
impact on heat flow. Consistently, we found in all of the examples that the
defects vanish with increased lead separation. Furthermore, in all cases the
thermal conductance of the defective chain was reduced in comparison to the
defect-free chain. The cumulated conductance revealed that no specific modes
within a particular energy range are responsible for the reduction. By
averaging over the set of geometries in the conductance-distance curve, we
estimated an average $\kappa_{\text{pn}}$ within a range of 20-30
pW/K. Finally, we calculated the room-temperature conductance within the
framework of NEMD. While we did not pull the junction, the use of large
reservoirs of gold atoms in the electrodes led to thermally induced
modulations of the electrode separation. These displacement fluctuations
caused a spontaneous appearance or vanishing of molecule-internal defects
within the simulation time. With the NEMD method we computed a thermal
conductance for C4-C10 in the range of 20-40 pW/K, which is consistent with
the previous DFT-NEGF approach.

We note that our NEMD simulations have important implications for using
molecule-internal disorder to tune the thermal conductance \cite{Li2015}. If
the time to measure the thermal conductance is faster than the transition
between different molecular structures, they will be identified in thermal
conductance measurements. In the opposite limit, the thermal conductance will
appear as an average over structures with straight or defective chains. This
latter case is the experimentally relevant situation, since measurements need
to average over many seconds to reveal a signal \cite{Cui2019}, while our
simulations show that gauche defects appear and vanish on picosecond time
scales. This kind of average due to a low time resolution of measurements was
considered in Ref.~\cite{Li2015} for graphene electrodes, which show a highest
phonon energy of around 185~meV as compared to the 20~meV of Au studied
here. The thermal average over different molecular configurations is expected
to result in a rather continuous mechanical tuning of the phonon heat
transport, and the corresponding increase in thermal conductance due to an
increasingly extended alkane would be overlaid by a weakly decreasing
radiative heat transport contribution with increasing electrode separation for
our three-dimensional metal electrodes \cite{Kloeckner2017a}.

In summary, our study reveals that the thermal conductance of molecular
junctions depends crucially on the strain state and is sensitive to the
precise atomic configuration. While the heat conductance value will depend on
the exchange-correlation functional or force field parametrization used in the
DFT-NEGF and NEMD approaches, the consistency of results is reassuring. It
confirms that anharmonic effects due to phonon-phonon scattering, taken into
account in the NEMD method, do not play a major role in these molecular
junctions containing only short molecules. In this sense we identify phonon
transport to be elastic and phase-coherent. Since the phonon transport through
single-molecule junctions is now experimentally accessible \cite{Cui2019}, we
hope to guide and stimulate future experiments with our study.

\section{Acknowledgments}
J.C.K.\ and F.P.\ thank P.\ Reddy, E.\ Meyhofer, J.\ C.\ Cuevas, D.\ M\"ohrle,
and A. Irmler for inspiring discussions on this project. In addition, both
authors gratefully acknowledge funding from the Carl Zeiss foundation, the
Junior Professorship Program of the Ministry of Science, Research and the Arts
of the state of Baden W\"urttemberg, and the Collaborative Research Center
(SFB) 767 of the German Research Foundation (DFG). An important part of the
numerical modeling was carried out on the computational resources of the bwHPC
program, namely the bwUniCluster and the JUSTUS HPC facility.


%

\end{document}